\documentclass{JINST}
\title{Status of the KM3NeT project}
\author{A. Margiotta$^a$ on behalf of the KM3NeT Collaboration\\
\llap{$^a$}INFN Sezione di Bologna and Dipartimento di Fisica e Astronomia - Universit\`a di Bologna,\\
viale C. Berti-Pichat, 6/2, 40127 - Bologna, Italy\\
E-mail: \email{annarita.margiotta@unibo.it}}
\abstract{KM3NeT is a deep-sea research infrastructure being constructed in the Mediterranean Sea. It will be installed at three sites: KM3NeT-Fr, offshore Toulon, France, KM3NeT-It, offshore Portopalo di Capo Passero, Sicily (Italy) and KM3NeT-Gr, offshore Pylos, Peloponnese, Greece. It will host the next generation Cherenkov neutrino telescope and nodes for a deep sea multidisciplinary observatory, providing oceanographers, marine biologists, and geophysicists with real time measurements. The neutrino telescope will search for Galactic and extra-Galactic sources of neutrinos, complementing IceCube in its field of view. The detector will have a modular structure and consists of six building blocks, each including about one hundred Detection Units (DUs). Each DU will be equipped with 18 multi-PMT digital optical modules. The first phase of construction has started and shore and deep-sea infrastructures hosting the future KM3NeT detector are being prepared in France near Toulon and in Italy, near Capo Passero in Sicily. The technological solutions for KM3NeT and the expected performance of the detector are presented and discussed.}
\begin{document}
\section{Introduction }\label{sec:intro}
The characteristics of the neutrino make it the perfect astronomical probe, able to cross extremely large distances and to transport information directly from the core of its production sites.
Until recently this idea was shown to be successful in two major occasions: when detectors on Earth intercepted neutrinos coming from the Sun~\cite{solar} and neutrinos produced during a SuperNova explosion in the Large Magellanic Cloud~\cite{magcloud}. In the first case the evidence of a tiny neutrino mass and of flavor oscillations was settled. The second measurement confirmed the basic nuclear processes of  stellar collapses. 
Recently, the first evidence of a high energy neutrino flux was announced by the IceCube Collaboration ~\cite{icecube}.  A bunch of neutrinos, whose energy ranges from 60 TeV to more than 1 PeV, from unresolved sources has been detected. This claim can be considered the opening of a new era for the exploration of the Universe, the age of neutrinos.
The same features making neutrinos so useful in the exploration of the far Universe and of dense astrophysical objects put severe constraints on the construction of an appropriate neutrino detector, requiring the instrumentation of huge amount of matter.
\section{Neutrino telescopes }\label{sec:nutel}
\begin{figure}[tbp] 
\hspace{3 cm}\includegraphics{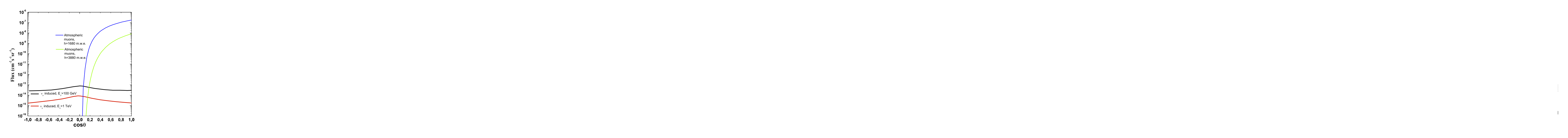}
\caption{Flux of $i)$ atmospheric muons (computed according to \cite{mupage0}) at two different depths and of $ii)$ atmospheric neutrino induced muons (from \cite{bartol}), for two different muon energy thresholds.}
\label{fig:muonspect}
\end{figure}

In the 60s Markov suggested the use of ocean or lake water as target and active medium for neutrino detectors~\cite{markov}. This idea is at the basis of the neutrino telescope concept. The neutrino telescope detection principle relies on the measurement of the Cherenkov light emitted in a natural transparent medium, like water or ice, along the path of charged  particles produced in  neutrino interactions in the vicinity of a three dimensional array of photon detectors. \\
Due to the long muon path length, the effective size of the detector is much larger for charged current  muon neutrino interactions than for other channels. Starting from time, position and amplitude of the photon signals, dedicated algorithms can reconstruct the trajectory of the muons, inferring the neutrino direction. The neutrino and the muon directions are almost collinear at high energy, and this allows the identification of a possible source with high resolution.
Measuring the total amount of light released in the detector, also the energy of the event can be evaluated within a factor 2-3.\\
Neutrino telescopes are located under large layers of matter, buried in the Antarctic ice or submersed under kilometers of sea water,  to stop the bulk of cosmic radiations.  Indeed, the flux of high energy atmospheric muons, produced in the interactions of primary cosmic rays with atmospheric nuclei, exceeds that of atmospheric neutrinos by several orders of magnitude, even at very large depth, see Fig.~\ref{fig:muonspect}. An efficient method to select neutrino-induced muons relies on the geometrical selection of reconstructed tracks, which rejects downward going muons due to atmospheric showers. 
A source of background that cannot be eliminated with simple geometrical  criteria is represented by atmospheric neutrinos.
A cosmic neutrino flux can be identified  either  by looking for an excess of events over the atmospheric neutrino background, or 
searching for a deformation of the atmospheric neutrino energy spectrum, typically due to an excess of events above a certain energy.
A search for coincident events with other astrophysical messengers provides a further possibility to reduce the background, looking for neutrino candidates at a specific time and from  known directions. This approach can be applied to transient sources like gamma ray bursts or micro quasars. \\
At present, an under-ice neutrino telescope, with an instrumented volume of about  1 km$^3$, is operated at South Pole, IceCube \cite {icecube-summ}. Underwater detectors have been built in the lake Baikal, in Russia, and in the Mediterranean Sea. Currently, the Baikal neutrino telescope is not taking data, due to an upgrade that will extend its volume to 1 Gton \cite {baikal}.
In the Mediterranean Sea, three collaborations, ANTARES  \cite {antares}, NEMO  \cite {nemo} and NESTOR  \cite {nestor}, have been active since a couple of decades.\\
The ANTARES collaboration built a 0.1 km$^3$ neutrino telescope close to the Provencal coast, in France. It is taking data in its full configuration since 2008. The ANTARES,  NEMO and NESTOR groups joined their efforts in long term R\&D activities creating the KM3NeT consortium, funded  by the EU in the framework of FP6 and FP7 in the period 2006-2012\footnote{FP6 Design Study; Project Acronym: KM3NET; Project Reference: 011937. FP7 Preparatory Phase; Project Acronym: KM3NET-PP; Project Reference: 212525}. In this context,  new technological solutions have been developed and deep sea sites were accurately characterized. At the beginning of 2013, the consortium has been transformed into a collaboration and started the construction of a KM3NeT Mediterranean neutrino telescope.
The KM3NeT Collaboration gathers more than two hundred scientists and engineers from ten European countries. Its main objective is the construction of a multidisciplinary observatory in the Mediterranean Sea, which will host a km$^3$-size  neutrino telescope,  complementary to  the IceCube detector in its field of view. In particular it will look directly to the Galactic centre. The KM3NeT infrastructure will include also a number of sea and Earth science  devices that will provide biologists, geophysicists, marine scientists with data collected at very large depth and will monitor the environmental conditions around the telescope.
\section{The physics case }\label{sec:phys}
More than one century after their discovery~\cite{hess-cr}, the origin and the acceleration mechanism of cosmic rays impinging on the Earth atmosphere have not been completely clarified yet. Primary cosmic rays are fully ionized atomic nuclei, mainly free protons.
\begin{figure}[tbp] 
\centering
\includegraphics[width=.4\textwidth]{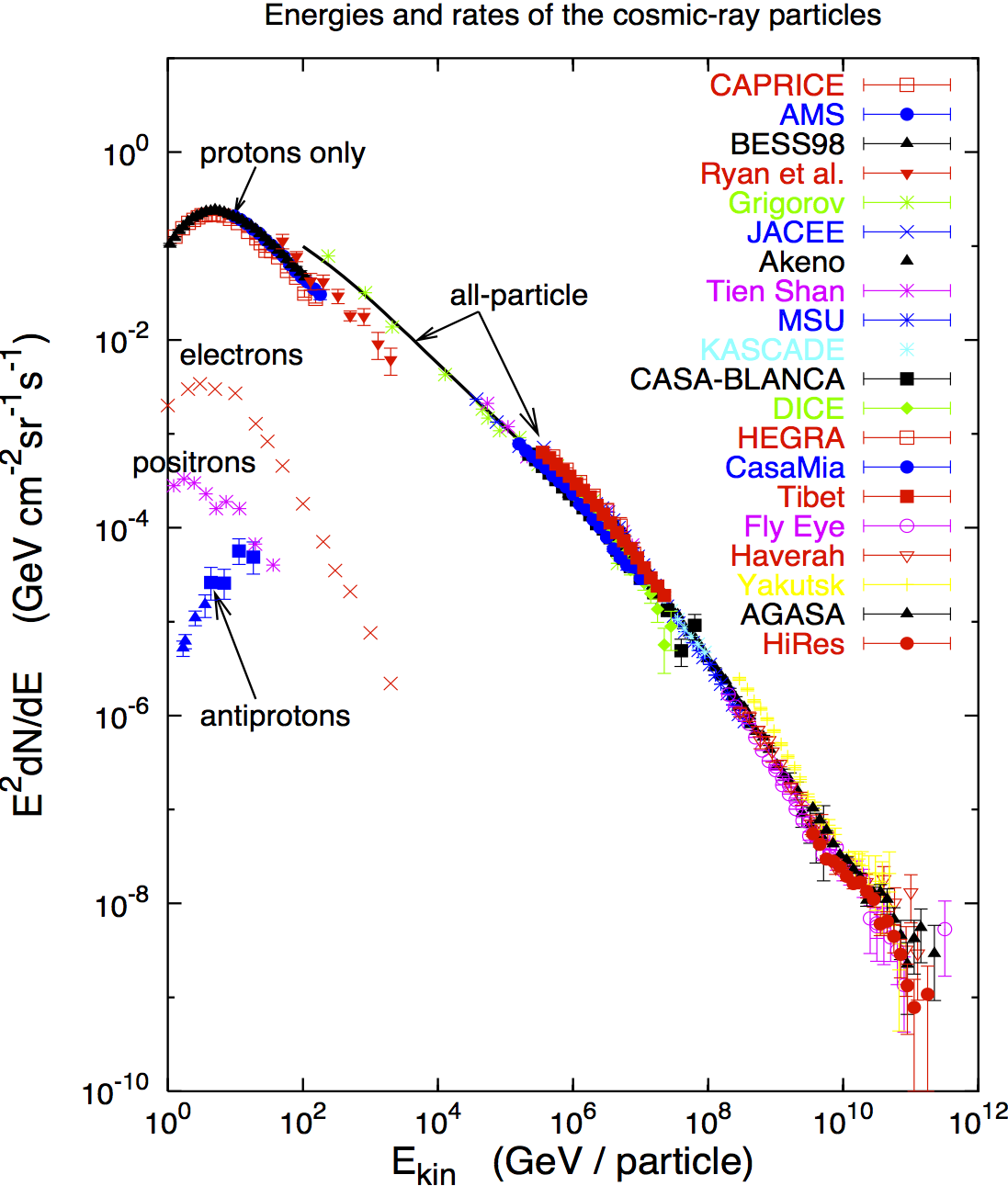}
\caption{Cosmic ray spectrum as measured on Earth. The contributions of protons, electrons, positrons and antiprotons at low energy, when measurements are available, are also reported. References to the experiments can be found in~\cite{crspec}. The figure is due to Tom Gaisser.}
\label{fig:crspectrum}
\end{figure}
Their energy spectrum spans over several decades from some MeV up to $10^{20}$ eV and falls steeply, showing some typical features connected to the specific production sites of the particles. The slope of the energy spectrum shown in Fig.~\ref{fig:crspectrum} changes near  $10^6$ GeV, the { \it knee}, and around $10^{10}$ GeV, the { \it ankle}.
Galactic Supernova remnants are generally considered as potential sites of production and acceleration of galactic CRs, below the { \it knee} energy.
More energetic CR cannot be explained with a Galactic origin.
The same astrophysical accelerators, Galactic and extra-Galactic, could be the responsible for production of TeV gamma rays detected by several ground based instruments~\cite{hess-magic} and for other extremely energetic phenomena like gamma ray bursts.\\
In the framework of a beam dump scenario, neutrinos can be produced in association with high energy gamma rays as the result of pion decay (hadronic model) when the accelerated hadrons, present in the core of massive objects, interact with ambient matter or photon fields:
{\small
\[
  p/A + A/\gamma \longrightarrow
\begin{array}[t]{l}
 \pi^0 \\
 \downarrow \\
 \gamma + \gamma
\end{array} +
\begin{array}[t]{l}
 \pi^\pm \\
 \downarrow \\
 \mu^\pm + \nu_\mu ~(\overline{\nu}_\mu) \\
 \downarrow \\
 e^\pm +\nu_e ~(\overline{\nu}_e) + \overline{\nu}_\mu ~(\nu_\mu)
\end{array} + N +...
\]}
Differently from what happens to other astrophysical messengers, they can deliver direct information on their production sites,  because they are not deflected by magnetic fields,  are stable and have a small interaction cross section.\\
Starting from the hypothesis of a common origin of neutrinos and gamma rays, benchmark fluxes have been evaluated and published in literature, which set the size of an astrophysical neutrino detector to the kilometre cube scale~\cite{vissani} \cite{wb}.\\
The main goal of the KM3NeT  telescope is the discovery of high-energy neutrino Galactic sources. Its location is optimal to look directly at the Galactic centre, which  hosts several TeV gamma-ray sources that, in a hadronic  scenario, could emit measurable neutrino fluxes. The excellent optical properties of water (scattering and absorption lengths) allow the reconstruction of muon tracks with an angular resolution better than one degree in the energy range where a signal is expected (a few TeV to a few 10 TeV). This makes the charged current interactions of muon neutrinos in the vicinity of the detector the golden channel for source identification.
Nevertheless, other channels, like neutral current interactions, will be explored as well. In the case of shower-like events, the angular resolution is expected to be around 10-15 degrees.\\
An intense work of simulation has been carried out in order to optimize the detector design to Galactic sources discovery.
In particular, the supernova remnant RXJ1713.7-3946 and the pulsar wind nebula Vela X have been studied, because gamma rays in the TeV region, which can be explained with the hypothesis of hadronic mechanisms acting inside these objects, have been measured coming from their directions.   Assuming an exponential cut-off power law for the neutrino energy spectrum and simulating different solutions for the detector design, the conclusion is that, with the detector arrangement described in section 4.2, KM3NeT can claim a discovery  after about 5 years of data taking, and the observation at a significance level of 3$\sigma$  with 50\%CL  after 2 years, for  RXJ1713.7-3946. A shorter data collection time (about 3 years for the discovery and slightly more than 1 year for the observation) would be necessary in the case of Vela X \cite{agata}.\\
Data from the  Fermi Large Area Telescope showed evidence of the emission of high energy gamma rays from two large areas, in the region above and below the Galactic Center, the so-called "Fermi bubbles"  ~\cite{fermi}. Assuming  a hadronic mechanism for the gamma ray production from the Fermi bubbles, a high energy neutrino flux is also expected. If  the gamma production is completely due to  hadronic processes, the results of Monte Carlo simulations ~\cite{fb-km3net} indicate that a discovery is possible in about one year of KM3NeT operation.\\
\section{Sites and technology }\label{sec:decision}
\subsection{Sites}\label{sec:sites}
Three sites have been identified as good locations for the KM3NeT, see Fig.~\ref{fig:sites}:
\begin{figure}[tbp] 
\centering
\includegraphics[width=.5\textwidth]{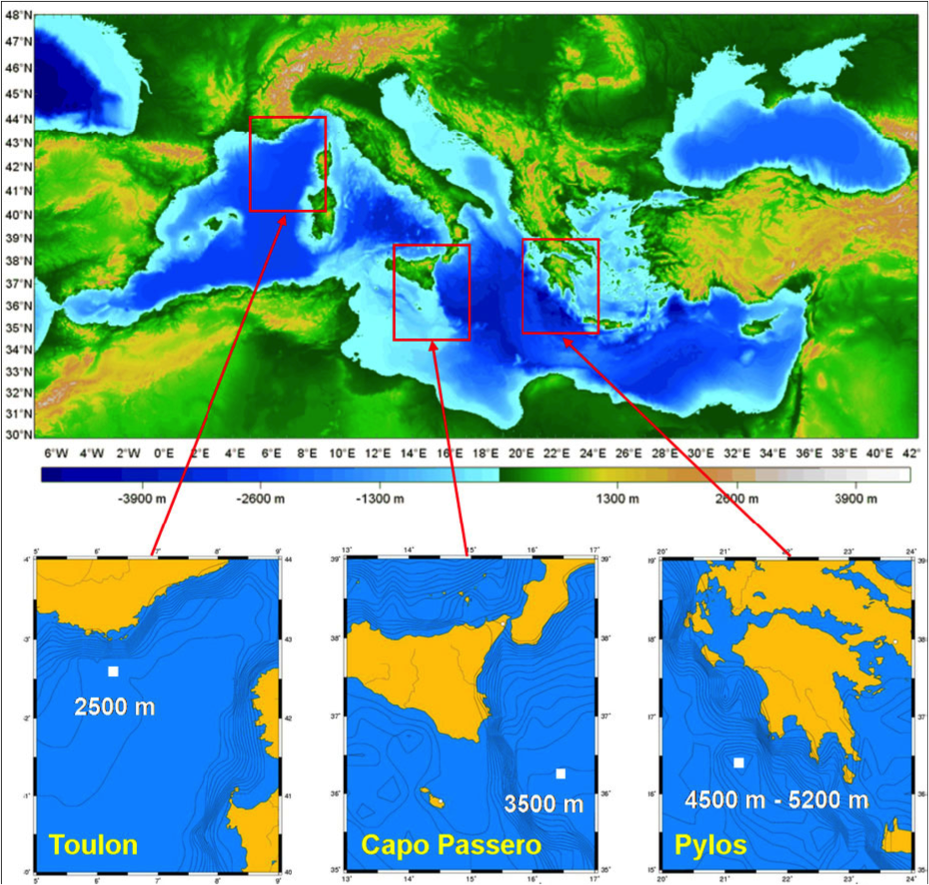}
\caption{Location of the three proposed sites for the construction of the KM3NeT neutrino telescope in the Mediterranean Sea.}
\label{fig:sites}
\end{figure}
\begin{itemize}
\item KM3NeT-Fr: this site is close to the ANTARES detector site, at a depth of about 2500 m, 20 km from the coast, south of Toulon;
\item KM3NeT-It: this is the site chosen by the NEMO Collaboration for the deployment of a prototype tower, which has been acquiring data since March 2013 \cite {tom}. It is at 3500 m under sea level at about 100 km from Capo Passero, in Sicily.
\item KM3NeT-Gr: this is the deepest site of the three proposed locations, at 4500/5000 m under sea level, at about 30 km southwest of Pylos, in Greece.  
\end{itemize}
Accurate simulations show that the global sensitivity to neutrino sources is not reduced if a distributed research infrastructure is built, provided each installation is large enough, at least half km$^3$. Such a distributed concept would also maximize the possibility of getting regional funding \cite {multisite}. \\
\subsection{Technology}\label{sec:tech}
\begin{figure}[tbp] 
\centering
\includegraphics[width=.2\textwidth]{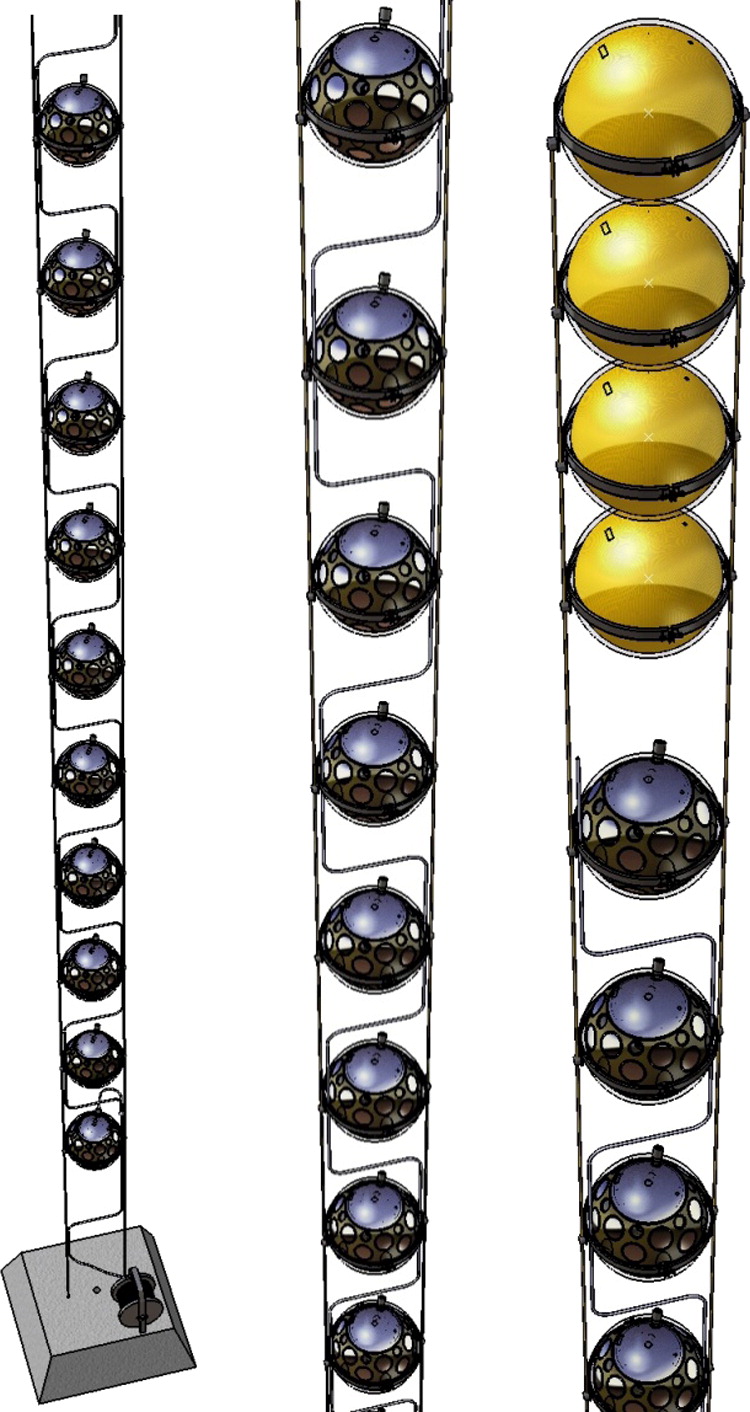}
\caption{Schematic view of a detection unit of KM3NeT.}
\label{fig:string}
\end{figure}

\begin{figure}[htb]
\centerline{
\includegraphics[width=0.4\columnwidth]{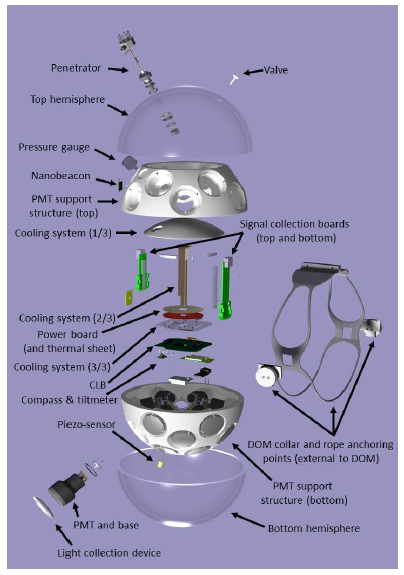}}
\caption{Internal structure of a  KM3NeTDOM.}
\label{fig:internal}
\end{figure}

\begin{figure}[tbp] 
\centering
\includegraphics[width=.4\textwidth]{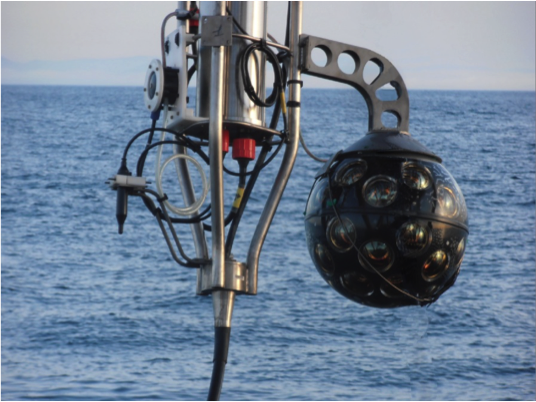}
\caption{Photo of the prototype multiPMT DOM, presently in data taking on the ANTARES detector.}
\label{fig:ppmdom}
\end{figure}

\begin{figure}[tbp] 
\centering
\includegraphics[width=.5\textwidth]{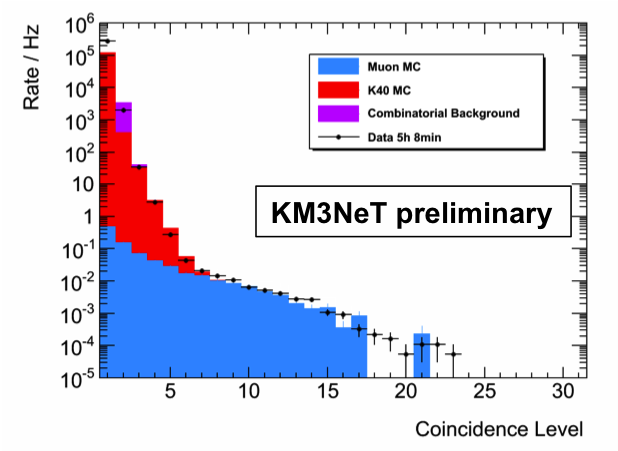}
\caption{Comparison between the measured and expected rate on the PPM-DOM hosted by the ANTARES detector as a function of the number of coincident hits. The red histogram represents the contribution due to the decay of $^{40}$K dissolved in the sea water. Measurements are still ongoing and the plot must be considered as preliminary.}
\label{fig:rate}
\end{figure}

\begin{figure}[htb]
\centerline{
\includegraphics[width=0.5\columnwidth]{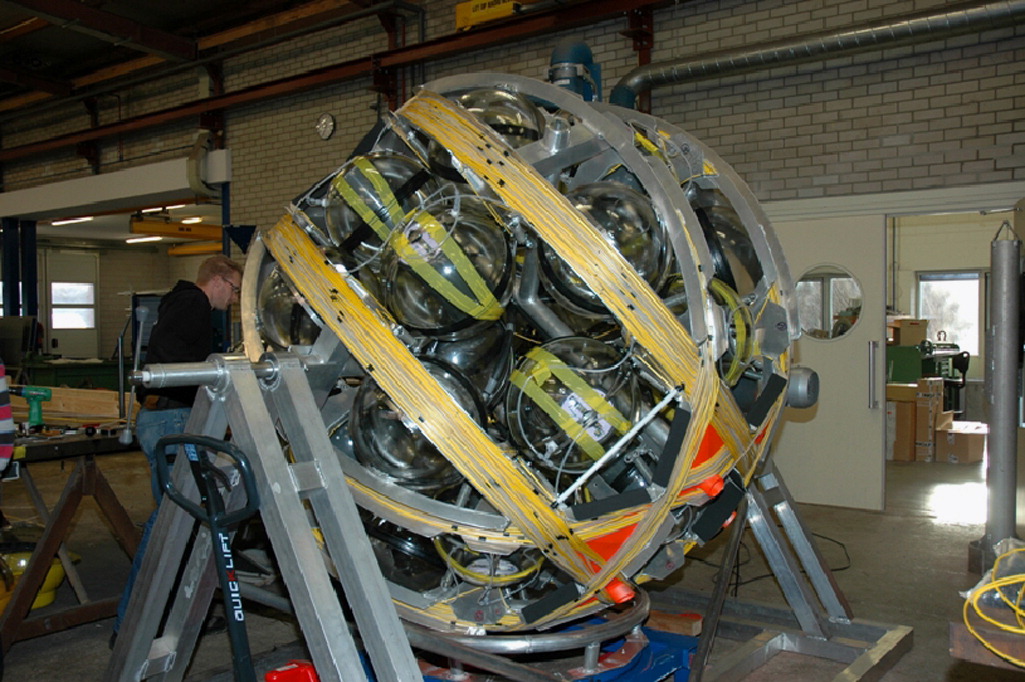}}
\caption{Photo of the launcher vehicle for optical modules (LOM) .}
\label{fig:lom}
\end{figure}

\begin{figure}[htb]
\centerline{
\includegraphics[width=0.3\columnwidth]{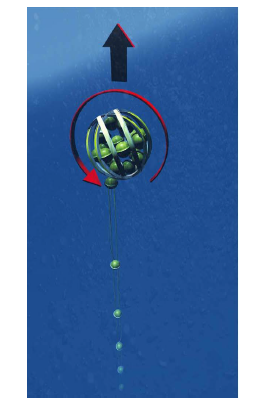}}
\caption{Unfurling a DU of KM3NeT.}
\label{fig:unroll}
\end{figure}

The  design of the detector is based on the concept of the Building Block. Each building block is made of 115 Detection Units (DU), at a distance of $\sim$ 90 m.\\
Each DU  consists of a flexible string, 700 m long, anchored at the sea bottom and kept taut by a system of buoys, carrying 18 storeys that will host one Digital Optical Module (DOM) each, with a vertical spacing of 36 m. The first storey will be at 100 m from the sea bottom. The mechanical structure is based on two Dyneema $^{\textregistered }$ ropes,  Fig.~\ref{fig:string}.\\
A Vertical Electro-Optical Cable (VEOC), an oil-filled pressure-balanced hose equipped with 18 optical fibers and two copper conductors,  connects all DOMs  to the DU base for power, readout and data transmission. \\
A major innovation concerns the design of the active part of a neutrino telescope, the optical module. The digital optical module used in KM3NeT is a 17-inch glass sphere, resistant to the high pressure present at the sea bottom,  housing 31 3-inch photomultiplier tubes (PMT)  and the active bases for power. The bases can be controlled individually from the shore in order to set the correct values of HV and threshold for each tube. A foam structure works as a support for the PMTs, with optical gel filling the cavity, to assure optical contact. 
A metallic structure glued on the sphere supports the electronic boards for their cooling, Fig.~\ref{fig:internal}. The maximum expected power consumption of a DOM is 10 W.\\
A multiPMT DOM has several advantages if compared to the traditional, large cathode single-PMT optical module used in ANTARES, Baikal and IceCube detectors, like, for example, a larger (three to four times) total photocathode area and a better discrimination of single vs multi photoelectron. A light collection ring further increases the effective photocathode area by 20-40\%. \\
\begin{table}[htdp]
\caption{Main characteristics required for the 3-inch PMT}
\begin{center}
\begin{tabular}[width=0.5\columnwidth]{lc}
\hline
Photocathode's diameter & $>$72 mm\\
Dynodes & 10 \\
Nominal Voltage for Gain $3\cdot 10^6$ & 900 1300 V \\
Gain slope & 6.5 min -8.0 max\\
QE at 404 nm &  $>$ 23 \%\\
QE at 470 nm & $>$ 18 \%\\
Collection efficiency & $>$ 87\% \\
Uniformity of QE and Coll. eff. & within 20\%\\
TTS (FWHM) & $<$ 5 ns \\
Dark count rate (0.3 s.p.e. threshold) &   $< $ 2 kHz\\
Pre-pulses &   $< $ 1 \% \\
Delayed pulses & $< $ 3.5 \% \\
Early afterpulses & $< $ 2 \% \\
Late afterpulses &  $< $ 10 \% \\
\hline
\end{tabular}
\end{center}
\label{default}
\end{table}
Three companies, ETEL, Hamamatsu and HZC, have developed new PMT types, according to the main characteristics required for the KM3NeT detector, see Table 1.
Some PMTs have been delivered for tests and evaluation, which are still in progress. \\
Each DOM acts independently as an IP node.  
All DOMs are synchronized to the subnanosecond level using a clock signal broadcast from shore. Monitoring and real-time correction for the propagation delays between the shore station and each single DOM will be performed using a White Rabbit application~\cite{wr}.
Also calibration sensors are included inside the optical modules: light beacons to illuminate groups of DOMs at known time, to monitor individual time offsets, piezo sensors for acoustic positioning, a tilt meter, a compass and sensors for temperature and humidity measurements inside the sphere.
The readout electronic boards (Central Logic Board, CLB) controls the data acquisition and communications with the shore station and are also hosted in the DOM.
The signal from a PMT  consists of the arrival time and the width of the pulse, measured as the time-over-threshold (ToT), typically set at 0.3 p.e., digitized and sent via a network of optical fibers to shore. Long range transmissions exploit DWDM techniques at 50 GHz spacing.  The "all-data-to-shore" concept is applied to the readout of the detector, following the experience of the ANTARES detector. On shore, the physics events are filtered from the background using a dedicated software and stored on disk.
The DAQ bandwidth is 200 Mb/s from each DOM and considers the possibility of  persistent bursting activity due to bioluminescent organisms.
The corresponding rate can vary depending on the site. More details are in ~\cite{tom}.
A fully equipped multiPMT-DOM has been installed on the ANTARES detector Fig.~\ref{fig:ppmdom} and has been taking data according to the expectations since spring 2013. Fig.~\ref{fig:rate} shows  a preliminary comparison between collected data and MC expectations. Simply requiring a coincidence among 6-7 PMTs   the bulk of hits due to the decay of $^{40}$K dissolved in the sea water can  be rejected. Measurements are still on going.
The deployment of such a large number of DU requires a special technique ~\cite{lom}. A recoverable launching vehicle has been designed and successfully tested during several cruises. Before deployment, each string is wound on a spherical aluminum launching vehicle, the LOM,  Fig.~\ref{fig:lom}. From the vessel, the LOM is lowered to the seabed, an acoustic release allows the launcher to freely float to the surface, where it can be easily recovered, while unrolling the string, Fig.~\ref{fig:unroll}. A Remotely Operated Vehicle (ROV) will connect each DU cable to a junction box linked to a sea floor infrastructure, receiving and delivering signal and power from/to the shore. Several qualification campaigns have been performed to validate the deployment concept. \\
\section{Present status of the project }\label{sec:status}
The first phase of the project (Phase-1) is funded mainly by Italian, Dutch and French contributions and foresees the construction and the deployment of 31 DUs, 24 DUs at the KM3NeT-It site and the remaining at the KM3NeT-Fr site. First strings will be deployed by early 2015.
The KM3NeT-It site will host also a set of 8 detection units constructed according to a previous design (NEMO tower, see  \cite {tom}), which will be operated independently.
The final complete configuration (Phase-2) foresees six building blocks corresponding to an instrumented volume of about 3 km$^3$, with each site hosting at least one building block.\\
Although KM3NeT Phase-1 and Phase-2 detectors have been optimised to the search for neutrinos sources, the recent results of the IceCube experiment  ~\cite{icecube} have prompted the KM3NeT Collaboration to consider also an intermediate KM3NeT Phase 1.5, whose main goal is to measure a diffuse flux neutrino signal, using different methodology, with complementary field of view and with a better angular resolution. The Phase1.5 infrastructure will comprise 2 building blocks to be realized at the KM3NeT-Fr and KM3NeT-It sites through an update of the existing installation. The additional estimated cost is about 50-60 MAC. Sensitivity studies are in progress.\\
\section{Conclusions}\label{sec:zzz}
The KM3NeT collaboration has just started to build a cubic kilometer  size neutrino telescope in the Mediterranean sea. It will complement the IceCube neutrino telescope, located at the South Pole, in its field of view. In particular, it will observe directly the Galactic Centre. The first phase of the project has  started with the aim of deploying  the first detection units by early 2015 at the Italian and French sites.
In the second phase of the project also the Greek site will be included. \\
Two major technological innovations will be used: the multiPMT digital optical module, which improves the rejection capability of the $^{40}$K background and increases the sensitive photocathode area, and the White Rabbit protocol, which will simplify the subnanosecond level of synchronization among DOMs.\\
An intense program of simulations has  started to define the optimal geometry and size for an early measurement of a diffuse flux of neutrinos and evaluate the KM3NeT sensitivity to the high-energy events recently published by IceCube.



%
%
%
%

\end{document}